\documentstyle[11pt,paspconf,psfig]{article}

\begin{document}

\title{Warm absorbers in Seyfert 1 galaxies}
\author{Christopher S. Reynolds}
\affil{JILA, University of Colorado, Boulder CO80309-0440, USA.}

\begin{abstract}
X-ray spectroscopy of Seyfert 1 galaxies often reveal absorption edges
resulting from photoionized gas along the line-of-sight to the central
engine, the so-called warm absorber.  I discuss how recent {\it ASCA}
observations of warm absorber variability in MCG$-$6-30-15 can lead us to
reject a one-zone model and, instead, have suggested a multi-zone warm
absorber.  The evidence for dust within the warm absorbers of MCG$-$6-30-15
and IRAS~13349+2438 is also addressed.  These dusty warm absorbers reveal
themselves by significantly reddening the optical flux without heavily
absorbing the soft X-ray photons.  Thermal emission from this warm/hot dust
may be responsible for the infra-red bump commonly seen in the broad band
spectrum of many Seyfert galaxies.
\end{abstract}

\section{Introduction}

X-ray observations of Seyfert 1 nuclei have provided direct evidence for
significant quantities of optically-thin photoionized gas along our
line-of-sight to the central X-ray source.  The first indications of such
material, which has become known as the {\it warm absorber}, were found with
low resolution X-ray spectroscopy of the luminous Seyfert nucleus
MR~2251-178 using {\it EXOSAT} (Halpern 1984).  The next significant
advancement came several years later when {\it ROSAT} PSPC
observations of the nearby Seyfert 1 galaxy MCG$-$6-30-15 revealed a broad
absorption feature at $\sim 0.8$\,keV (Nandra \& Pounds 1992).  This was
interpreted as being a blend of the photoelectric K-shell edges of {\sc
O\,vii} and {\sc O\,viii} (at 0.74\,keV and 0.87\,keV, respectively) which
are imprinted upon the primary X-ray spectrum when the line-of-sight to the
central engine passes through the warm absorber.

The CCDs on board the US/Japanese X-ray satellite {\it ASCA} provide an
order of magnitude improvement in spectral resolution over the {\it ROSAT}
PSPC and allow us to study warm absorbers in detail.  {\it ASCA}
observations can separate and independently measure the properties of the
{\sc O\,vii}/{\sc O\,viii} edges.  Detailed scrutiny of the {\it ASCA}
spectra of bright Seyfert 1's can also reveal the presence of other edges
and lines, although oxygen edges usually dominate.  {\it ASCA} has shown
us that approximately half of all Seyfert 1 nuclei display a warm absorber
(Reynolds 1997).

In this contribution, I shall describe two studies that address the physical
nature of the warm absorber.  Due to the fact that it has been most
extensively observed by {\it ASCA}, I will focus on results for the Seyfert
1 galaxy MCG$-$6-30-15 ($z=0.008$).

\section{Warm absorber variability}

During Performance Verification observations, {\it ASCA} found large
variations in the depths of the oxygen edges in MCG$-$6-30-15 (Fabian et
al. 1994; Reynolds et al. 1995).  The timescales of this variability were
found to be as short as $\sim 10\,000$\,s.  However, the physical nature of
this variability was unclear.  Within the context of a one-zone
photoionization model (characterized by the column density $N_{\rm W}$ and
the ionization parameter $\xi$), a drop in the primary X-ray flux was
accompanied by an increase in both the column density and ionization
parameter.  It is difficult to imagine a situation in which the total
absorber column density is correlated to the intrinsic X-ray flux on such
short timescales (Reynolds et al. 1995).

\begin{figure}
\hspace*{1cm}
\psfig{figure=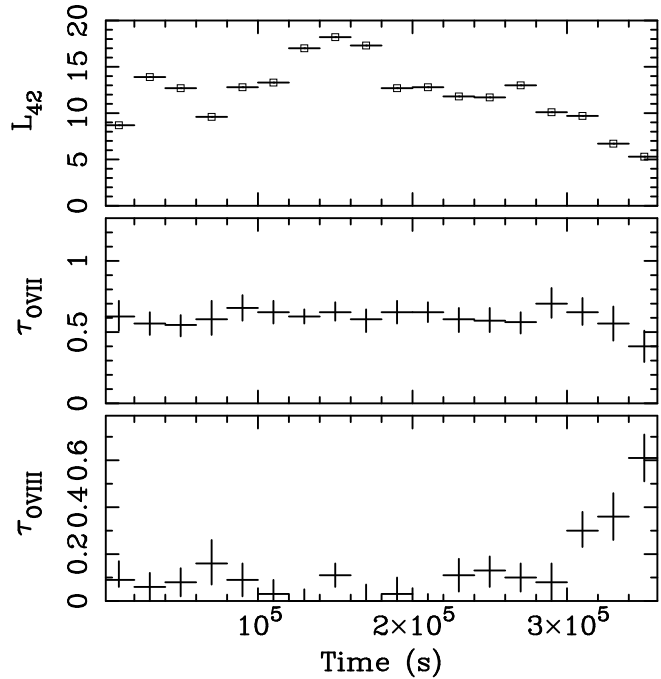,width=1.0\textwidth,angle=270}
\caption{Intrinsic 2--10\,keV luminosity and the warm oxygen edge depths as a
function of time for the long {\it ASCA} observation of MCG$-$6-30-15.
This assumes $H_0=50\,{\rm km}\,{\rm s}^{-1}\,{\rm Mpc}^{-1}$.}
\end{figure}

A long {\it ASCA} observation of MCG$-$6-30-15 resolved this confusion.
Figure~1 shows the intrinsic X-ray luminosity and the warm oxygen edge
depths as a function of time for this observation.  It can be seen that the
{\sc O\,vii} edge depth remains essentially constant whereas the {\sc
O\,viii} edge depth is strongly anti-correlated with the primary X-ray flux
(Otani et al. 1996).  The fact that the warm absorber variability is
directly related to the primary X-ray variability is direct evidence that
the warm absorber is photoionized.  However, the constancy of the {\sc
O\,vii} edge is inconsistent with a one-zone absorber model.  To see this,
we suppose that the warm absorber can be described as a single uniform slab
and note that the recombination and photoionization timescales of this
material must be short in order to see any variability of the edge depths.
The relative depths of the oxygen edges fixes the ionization parameter of
this slab to be $\xi\sim 20\,{\rm erg}\,{\rm cm}\,{\rm s}^{-1}$.  However,
for this value of $\xi$, photoionization models clearly show that the {\sc
O\,vii} edge depth should decrease, and the {\sc O\,viii} edge depth should
increase, as the primary ionizing flux increases.  This is contrary to
observations.

\begin{figure}[t]
\psfig{figure=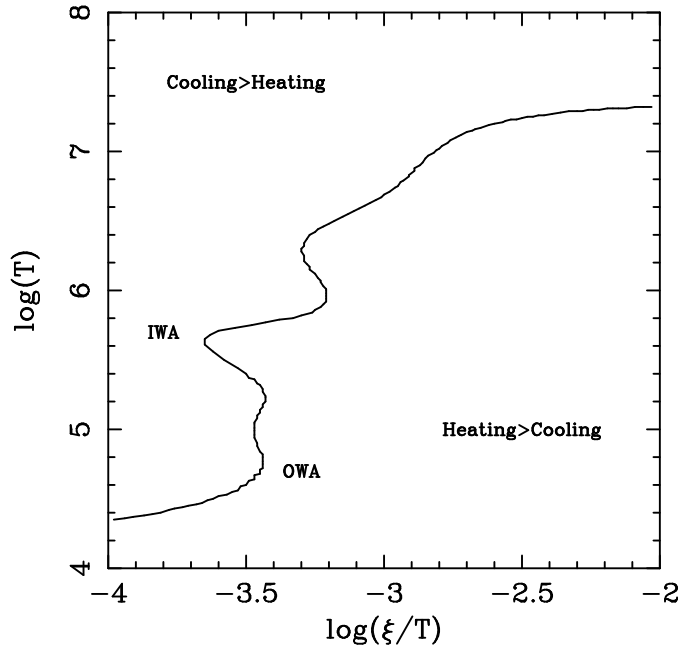,width=1.0\textwidth,angle=270}
\caption{Thermal equilibrium curve for photoionized plasma.   The assumed
ionizing continuum consists of a single power-law with photon index
$\Gamma=1.8$ extending from 13.6\,eV to 40\,keV.   The positions of the
inner warm absorber (IWA) and outer warm absorber (OWA) are shown.}
\end{figure}

The simplest extension of the one-zone model, a two-zone model, can nicely
account for the observed behaviour (Otani et al. 1996).  In this model,
much of the {\sc O\,vii} edge results from a distant, tenuous warm absorber
which has a long recombination timescale and so is not seen to react to
changes in the ionizing flux.  Spectral fitting shows this absorber to have
a column density of $N_{\rm W}\approx 5\times 10^{21}\,{\rm cm}^{-2}$, an
ionization parameter of $\xi\approx17$, and to be at a distance of
$R>1$\,pc from the central engine.  By contrast, much of the {\sc O\,viii}
edge originates from an inner warm absorber which is highly ionized and has
a short recombination timescale.  A drop in the ionizing flux leads to
recombination of fully-stripped oxygen into {\sc O\,viii} ions and thus an
increase in the {\sc O\,viii} edge depth.  This region is constrained to
have $N_{\rm W}\approx 1\times 10^{22}\,{\rm cm}^{-2}$, $\xi\approx 70$ and
$R<10^{17}$\,cm.  It is tempting to identify this component with an
optically-thin component of the broad line region.

It is important to account for the two-zone nature of this absorber when
discussing the thermal stability of the warm material.  The thermal
stability of photoionized plasma to isobaric perturbations can be studied
by examining the thermal equilibrium curve on the ($\xi/T$, $T$) plane
(McCray 1979; Krolik, McKee and Tarter 1981).  Parts of this curve which
have negative slope and are associated with a multi-valued regime
correspond to thermally unstable equilibria.  A one-zone model for the warm
absorber implies that it is either thermally unstable or exists in
extremely small pockets of stable parameter space (Reynolds \& Fabian
1995).  However, given a two-zone parameterization, the situation is rather
different, as shown in Fig.~2.  We find that the outer warm absorber can be
understood as material at the extreme (hot) end of the stable cold branch,
whereas the inner warm absorber is material at the extreme (cold) end of
the stable intermediate/hot branch.  Whilst the exact form of this curve is
sensitive to the ionizing continuum shape assumed, it is clear that the
two-zone nature of the absorber must be accounted for in any discussion of
thermal stability.

\section{Optical/UV extinction and dusty warm absorbers}

Whilst reporting the results of the first optical spectroscopy of
MCG$-$6-30-15, Pineda et al. (1980) noted that the X-ray spectrum does {\it
not} display the neutral absorption expected on the basis of its optical
reddening.  This issue has recently been investigated in more detail by
Reynolds et al. (1997).  Both the optical emission line spectrum and the
continuum form suggest that this Seyfert nucleus is reddened by $E(B-V)\sim
0.6$ or more.  Assuming a standard LMC extinction law and using a Galactic
dust/gas ratio implies that a gas column of at least $N_{\rm H}\sim 3\times
10^{21}\,{\rm cm}^{-2}$ should accompany this reddening.  However, X-ray
spectroscopy with both {\it ASCA} and {\it ROSAT} can set stringent limits
on the cold (i.e., neutral) gas column of $N_{\rm H}<2\times 10^{20}\,{\rm
cm}^{-2}$.

A similar situation has been found for the quasar IRAS~13349+2438 (Brandt
et al. 1996).  In this object, spectropolarimetry reveals a heavily
reddened direct continuum and an unreddened scattered component (Wills et
al. 1992).  X-ray observations reveal a variable X-ray source (indicating
that we are viewing the X-ray source directly, rather than through
scattered flux) with a warm absorber but no detectable neutral absorption.
The X-ray upper limits on the neutral column are an order of magnitude
lower than that expected on the basis of the reddening (Brandt et al. 1996,
1997).

Since the X-ray flux is thought to originate from deeper within the central
engine than the optical flux, this discrepancy is difficult to understand
as a purely geometric effect.  A dust/gas ratio which is 10 times the
Galactic value would reconcile these studies, but is difficult to
understand physically (we might expect dust grains to be preferentially
destroyed near an AGN, but not preferentially created).  The most
reasonable resolution of this problem is that the dust responsible for the
optical reddening resides in the warm absorber.  In both MCG$-$6-30-15 and
IRAS~13349+2438, the inferred dust/warm-gas ratio is similar to the
Galactic dust/cold-gas ratio.  In other words, the warm absorber appears to
represent Galactic-like dusty gas that has been ionized with little
destruction of the dust grains.

Under the conditions envisaged, the dust grains are thermally decoupled
from the relatively tenuous gas.  They come into a thermal equilibrium such
that the thermal radiation emitted by each grain balances the incident AGN
flux on that grain.  Grains can survive provided two conditions are
satisfied.  First, the grains must be sufficiently far from the central
engine so that they do not sublime.  In the case of MCG$-$6-30-15, this
translates into a distance limit of $R>10^{17}$\,cm and suggests that it is
the outer warm absorber, as opposed to the inner warm absorber, which is
dusty.  Secondly, the gas temperature cannot exceed $10^6$\,K or else the
grains would be destroyed via thermal sputtering.  This is easily satisfied
if the warm absorber is photoionized rather than collisionally ionized.
This dust may be responsible for a significant part of the infra-red bump
seen in the broad-band spectrum of many Seyfert nuclei.

\section{Conclusions}

Two studies relevant to the physical nature of the warm absorber have been
described.   To summarize the results of these investigations:
\begin{enumerate}
\item The temporal variability of the warm absorber in MCG$-$6-30-15 as
seen by {\it ASCA} suggests that it is comprised of at least two zones.
The inner zone may be related to the broad line region whereas the outer
zone is at radii characteristic of the putative molecular torus and the
narrow line region.
\item The absence of neutral absorption, which is naively expected to
accompany the optical reddening in both MCG$-$6-30-15 and IRAS~13349+2438,
suggests that the warm absorber may be dusty.  In MCG$-$6-30-15, this dust
must lie in the outer warm absorber or else it would be sublimated by the
intense AGN radiation field.
\end{enumerate}

\acknowledgments

I am indebted to Andy Fabian, Chiko Otani and Martin Ward with whom much of
this work was performed in collaboration.  This work was supported by a
PPARC (UK) studentship and the National Science Foundation under grant
AST-9529175.

\begin{question}{N.~Arav}
How well can you constrain the properties of the two-zone model for
explaining the {\sc O\,vii}/{\sc O\,viii} edges?
\end{question}
\begin{answer}{C.~Reynolds}
MCG$-$6-30-15 shows a fairly clean pattern of variability -- the {\sc
O\,vii} edge remains  essentially constant whilst only the {\sc O\,viii}
edge varies.   In this case we can decouple the two zones quite easily.
In a more general case (i.e., both oxygen edges varying) it is rather more
difficult.   This may be the case in some other Seyfert galaxies (e.g.,
NGC~3227) in which a complex pattern of warm absorber variability is seen.
\end{answer}

\begin{question}{C.~Foltz}
What is known about the ultraviolet spectrum of MCG$-$6-30-15?  Is Mg\,{\sc
ii} absorption seen?
\end{question}
\begin{answer}{C.~Reynolds}
Not much is known.  {\it IUE} shows it to be a rather faint UV source and
I'm not aware of a {\it HST} UV spectrum of this object.  To my knowledge,
the only UV feature that has definitely been detected is the {\sc
C\,iv}$\lambda 1549$ line (even the continuum is barely detected).  This is
all consistent with it being heavily reddened.
\end{answer}

\begin{question}{E.~Agol}
What happens to the Fe K$\alpha$ line as the warm absorber varies?  Is
there any evidence for iron edge absorption?
\end{question}
\begin{answer}{C.~Reynolds}
As the X-ray flux is seen to enter a low state, the warm absorber is seen
to change in the manner I described, and the iron line is seen to get
broader and stronger.  However, the iron line is thought to originate from
the innermost regions of the accretion flow and so is probably not
directly related to any changes in the warm absorber.  {\it ASCA} does not
detect any iron edge, but the constraints aren't very strong due to the
limited sensitivity of {\it ASCA} at energies above the iron line.
\end{answer}

\end{document}